# Wiki-MetaSemantik: A Wikipedia-derived Query Expansion Approach based on Network Properties

D. Puspitaningrum[1], G. Yulianti[2], I.S.W.B. Prasetya[3]
[1,2]Department of Computer Science, The University of Bengkulu
WR Supratman St., Kandang Limun, Bengkulu 38371, Indonesia
[3]Department of Information and Computing Sciences, Utrecht University
PO Box 80.089, 3508 TB Utrecht, The Netherlands
E-mails: diyahpuspitaningrum@gmail.com[1], gries.cs@unib.ac.id[2], s.w.b.prasetya@cs.uu.nl[3]

*Abstract*- **This paper discusses the use of Wikipedia for building semantic ontologies to do Query Expansion (QE) in order to improve the search results of search engines. In this technique, selecting related Wikipedia concepts becomes important. We propose the use of network properties (degree, closeness, and pageRank) to build an ontology graph of user query concept which is derived directly from Wikipedia structures. The resulting expansion system is called Wiki-MetaSemantik. We tested this system against other online thesauruses and ontology based QE in both individual and meta-search engines setups. Despite that our system has to build a Wikipedia ontology graph in order to do its work, the technique turns out to works very fast (1:281) compared to other ontology QE baseline (Persian Wikipedia ontology QE). It has thus the potential to be utilized online. Furthermore, it shows significant improvement in accuracy. Wiki-MetaSemantik shows better performance in a meta-search engine (MSE) set up rather than in an individual search engine set up.**

## I. INTRODUCTION

Wikipedia is the largest human-built knowledge repository currently in existence, available in over 250 languages, with characteristics such as: it is not limited in scale, includes dense link structures, URL-based word sense disambiguation, and brief anchor [13][7]. Wikipedia's structures consists of: articles, disambiguation pages, redirects, hyperlinks, category structure, templates and infoboxes, discussion pages, and edit histories. An article describes a single concept. Each article's title resemble terms in a conventional thesaurus. Terms of similar meaning are linked to an article using redirect. The disambiguation pages allow users to select an article. Hyperlinks in an article express relationships to other articles. All of these structures form a various number of concepts of human knowledge that can be exploited as a tool to build a semantic ontology of any concept.

Since currently available Wikipedia ontologies only map small parts of human knowledge concepts, in this paper, we propose a new method of query expansion (QE), called Wiki-MetaSemantik, to build an ontology automatically from Wikipedia. Wiki-MetaSemantik exploits all the benefits of Wikipedia structures and alleviates topic drift that otherwise often appears when using synonyms of online thesauruses such as WikiSynonyms, WordNet, and Moby. Existing Pseudo-Relevance Feedback (PRF) query expansions suffer from several drawbacks such as query-topic drift [8][10] and inefficiency [21]. Al-Shboul and Myaeng [1] proposed a technique to alleviate topic drift caused by words ambiguity and synonymous uses of words by utilizing semantic annotations in Wikipedia pages, and enrich queries with context disambiguating phrases. Also, in order to avoid expansion of mistranslated words, a query expansion method using link texts of a Wikipedia page has been proposed [9]. Furthermore, since not all hyperlinks are helpful for QE task though (e.g. a link text "French" inside a page titled "baseball" is not very helpful to expand a query on the latter), Farhoodi et al. [3] proposed a QE method using ontology derived from Wikipedia, or the Wikipedia Persian ontology method for short. They used weights to capture relationships between Wikipedia structures.

Our Wiki-MetaSemantik selects only one most relevant Wikipedia knowledge concept of a user's query and generates QE using combinations of degree, closeness, and pageRank of the ontology graph built by the chosen knowledge concept. In selecting QE, terms distributions and structures of Wikipedia pages are taken into account. We compare the precision and scalability of the proposed method against an existing QE method which also use an ontology derived from Wikipedia. Thirty multi domain queries are used in this comparison. Also, since combining multiple datasets can lead into better accuracy [14], for example as a meta-search engine [18], we propose the use of meta-pseudo relevance feedback (meta-PRF) for automatic judgement purpose as in [17][18]. Our experiments show benefit of IR performance when using Wiki-MetaSemantik.

The main contribution of this work is that our approach, Wiki-MetaSemantik, potential for fast relevant building of Wikipedia ontology from query of any concept. Despite its modesty, we believe the algorithm to be potential for running on the fly thus it can be embedable in either a meta-search or individual search engine. By building an ontology online, our algorithm does not really depend on how extensive, or limited, related concepts are documented in Wikipedia. Hence, the algorithm is more flexible in capturing the semantic of any user query.

The remainder of this paper is organized as follows. Section 2 discusses related research on QE. Section 3 describes our proposed methods for using ontology-derived from Wikipedia as meta-pseudo relevant documents. Experimental results are reported in Section 4. We summarizes the outcomes, possible limitations, and the future work directions in Section 5.

## II. Related Work

Hsu et al. [4] shows the use of WordNet in query expansion. Performance of queries expanded by WordNet outperforms that of queries without expansion, and queries expanded with a single resource. Semantic graphs are commonly used to model word senses and are usually built using thesauri or lexical databases such as WordNet [19]. Approaches such as PageRank, HITS or node similarity can be used to second alternative queries [11][20]. Bruce et al. [2] uses Wikipedia and its hyperlink structures to find related terms for reformulating a query using link probability weighting and link based measure.

The work by Farhoodi et al. built query expansion for Persian ontology [3]. To improve the results of retrievals, they proposed to exploit the following: 1) the relation between title and keywords, 2) the relation between the title and the concepts in article's text, and 3) the relation between the title and the concepts in 'See also' links. These relations are given the weights of 0.6, 0.5, and 0.7 respectively. Most of the results have higher precision when the query expansion is implemented but the precision may fall depending on the quality of Wikipedia pages and the links in these pages.

Our query expansion method takes advantages of the above mentioned work [3], by reusing the set up weights of Wikipedia relations. However, the method itself works in very differently. Assume we have a graph/network of a small world (e.g. a set of selected Wikipedia concepts). A user query becomes the root node and structures such as titles, keywords, text, 'See also', and 'Category' become leafs or nodes in an ontology graph, up to a certain number of hops of Wikipedia pages. Each node has links to other Wikipedia pages, which become the edges in the graph. QE terms are generated using a carefully set up weights and combinations of graph-based measures (degree, closeness, and pageRank).

*Graph centrality measures* are used to determine how important a node is in a network/graph [15]. In *degree centrality*, a node is important if it has many edges connected to and from it. *Network degree* is the maximum degree centrality over a network's nodes. In *closeness centrality*, a node is important if it is "close" to all other nodes in the network, in terms of the sum of the shortest paths to all other nodes [5]. In PageRank, a term is as valuable as other terms that link to it. According to Page et al. [16], a typical analogy for this is that a link from one page to another essentially can be seen as a vote being cast by one page onto another. In our query expansion case, a node is voted by the number of its backlinks (the links from other nodes to that node).

## III. Wiki-MetaSemantik: The Development

Wiki-MetaSemantik search engine system is divided into eight major steps. Fig. 1 explains these eight steps.

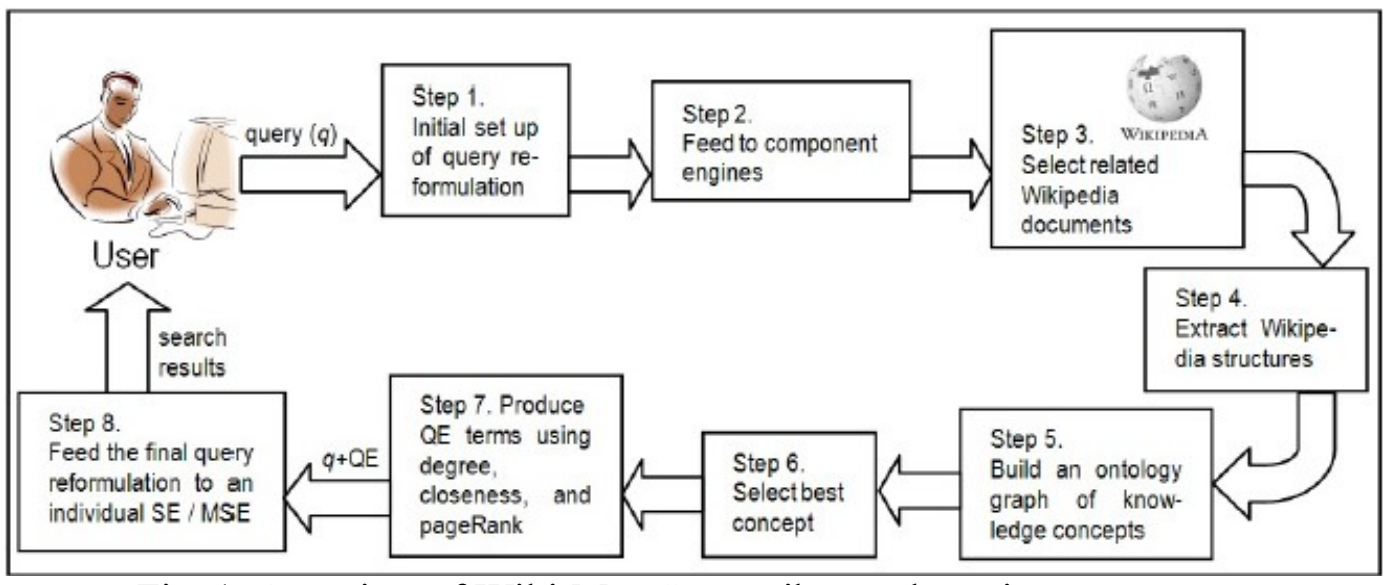

Fig. 1. Overview of Wiki-MetaSemantik search engine system

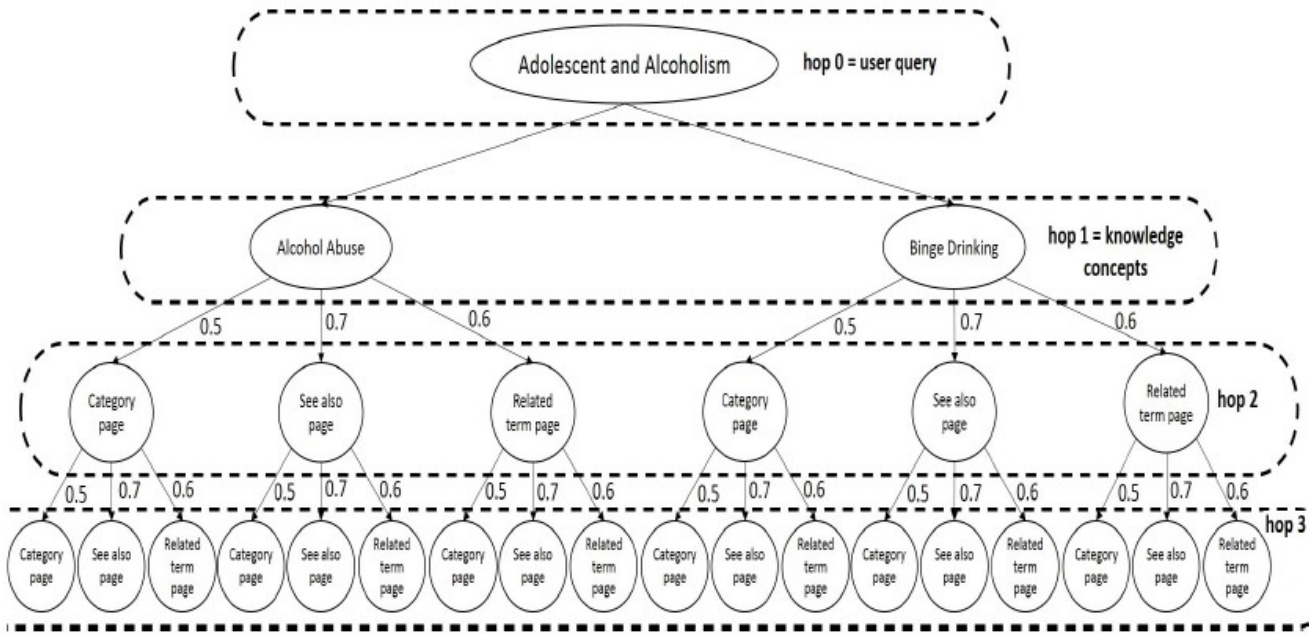

Fig. 2. A typical or modest example of an ontology graph from query "*adolescent and alcoholism*" (truncated from original due to space)

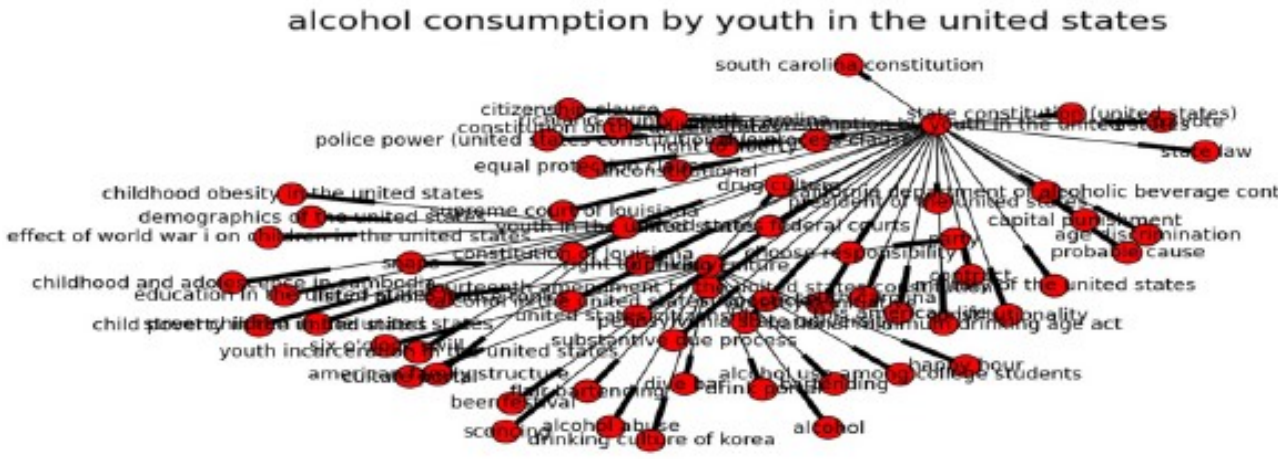

Fig. 3. First Hop of "*alcohol consumption by youth in the united states*" concept

**Step 1: Initial set up of query reformulation**. For each query, we add it with "*wikipedia*" word. For example query "*adolescent and alcoholism*" has initial query format of "*adolescent and alcoholism wikipedia*".

**Step 2: Feed the initial query reformulation to component engines.** Feed the initial query to each component engines of a meta-search engine, or to an individual one.

**Step 3: Select related Wikipedia documents.** Choose only top-documents originated from any Wikipedia domains.

**Step 4: Extract Wikipedia structures.** Visit each Wikipedia pages and extract related keywords from the pages by exploiting all Wikipedia structures: Title, 'Category', 'See also', and 'Related terms'. 'Related terms' denotes all terms in the Wikipedia passages that marked by blue fonts (hyperlinks, or assumed as relevant regarding to the corresponding Title).

Each initial Wikipedia pages (hop 1), as a result of Step 3, is a candidate of knowledge concepts of the user query. For instance user query "*adolescent and alcoholism*" results 8 candidates of knowledge concepts: 1) Alcoholism (https://en.wikipedia.org/wiki/Alcoholism), 2) Alcoholism in adolescence (https://en.wikipedia.org/wiki/Alcoholism_in_adolescence), 3) Alcohol abuse (https://en.wikipedia.org/wiki/Alcohol_abuse), 4) Binge drinking (https://en.wikipedia.org/

wiki/binge_drinking), 5) Alcohol consumption by youth in the United States-Wikipedia (https://en.wikipedia.org/wiki/alcohol_consumption_by_youth_in_the_United_States), 6) Alcoholism in family systems (https://en.wikipedia.org/wiki/Alcoholism_in_family_systems), 7) Substance abuse (https://en.wikipedia.org/wiki/Substance_abuse), 8) Alcohol and health (https://en.wikipedia.org/wiki/alcohol_and_health).

**Step 5: Build an ontology graph of knowledge concepts**. Do Step 4 to every existing Wikipedia hyperlinks until 3rd hop, and then create an ontology graph (Fig. 2). Fig. 3 shows an ontology graph generated from a part of nodes of user query "*adolescent and alcoholism*" which represented by "*alcohol consumption by youth in the united states*" concept.

**Step 6: Select best concept.** Isolate the ontology graph into small graphs, starts by taking candidates of knowledge concepts as roots of each those small graphs and then respectively go down to all descendants (or all related nodes). Choose one of the isolated graphs with the highest degree as the best concept. Degree of graph shows a graph importance.

**Step 7: Produce QE terms using degree, closeness, and pageRank.** For all nodes in the best concept, compute their degree, closeness, and pageRank scores and then create lists of nodes in decrease order sorted separately by their degree, closeness, and pageRank scores. After that find sets of nodes (or keywords) that appear in degree, closeness, and pageRank lists. Different way of intersect the lists will produce different keywords and order of keywords. There are 3 ways of intersect those lists:

1) Intersection_set#1: Find 100-top nodes from degree list that also appear in closeness list and pageRank list;
2) Intersection_set#2: Find 100-top nodes from closeness list that also appear in degree list and pageRank list;
3) Intersection_set#3: Find 100-top nodes from pageRank list that also appear in closeness list and degree list.

The 3 lists then combine into one list using Borda Count voting technique. Post-processing technique then must be done by applying filter that taking only terms that are neither in user query nor stopwords. Final QE terms, or just QE terms for simplicity, are taken from the remain Borda Count list for top-2 or top-3 terms depend on user needs.

**Step 8: Feed the final query reformulation to an individual search engine or to a meta-search engine.** The final query reformulation is defined by: user query + QE terms. Fetch them to an individual or to a meta-search engine (MSE) along with graph properties weights of degree, closeness, and pageRank to get search results. The weights are input parameters to create an MSE.

## IV. Experimental Evaluation

### A. Test Data

Table I shows the basic queries. They are expanded to 30 multi domain queries using combinations of operator 'AND/OR'. We use 4 baseline methods compare to the Wiki-MetaSemantik. Method 1 to 3 use synonyms from online thesauruses (WordNet, Moby Thesaurus, or WikiSynonyms respectively), Method 4 use QE using the Wikipedia Persian ontology method.

TABLE I
The Basic Multi Domain Queries [17][12]

| Two Terms | Three Terms |
|---|---|
| database overlap | comparative education methodology |
| multilingual OPACs | java applet programming |
| programming algorithm | indexing digital libraries |
| roadmap plan | geographical stroke incidence |
| adolescent alcoholism | culturally responsive teaching |

We did all of our experiments on either individual search engines or meta-search engines from 5 popular search engines: Google, Bing, Lycos, Ask, and Exalead. For the meta-search engines, we use 10 combinations of three component search engines: (Google-Lycos-Bing), (Google-Lycos-Ask), (Google-Lycos-Exalead), (Google-Bing-Ask), (Google-Bing-Exalead), (Google-Ask-Exalead), (Lycos-Bing-Ask), (Lycos-Bing-Exalead), (Lycos-Ask-Exalead), and (Bing-Ask-Exalead).

All experiments were tested in a laptop with a 2 GB processor, a 80 GB hard disk, 2 GB memory, modem UMTS 850/1900/2100 MHz 7.2 Mbps. The Wiki-MetaSemantik is built using Python version>2.7.9, running on Windows 10 with additional Python modules: NetworkX, BeautifulSoup, NLTK, Mechanize, Wikipedia, and WebPy.

### B. Measurements

*Baselines*

For evaluation, we compare our Wiki-MetaSemantik method against other QE methods: the synonyms or thesaurus-based methods (WordNet, WikiSynonyms, Moby Thesaurus), and the ontology-based method (the Wikipedia Persian Ontology QE, [3]). The WordNet and the WikiSynonyms QE methods work by search over top-*k* synonyms in either WordNet or WikiSynonyms. The Moby thesaurus QE methods works by search over *k* random synonyms in the Moby Thesaurus. Both WordNet and WikiSynonyms are in decrease order by synonymity, whereas Moby are in equal weight of synonymity.

In the Wikipedia Persian Ontology QE, the existent concepts in the query are mapped on to the ontology graph based on Wikipedia relationship. The ontology is then used to expand user queries and submitted to the search engine to get the search results. Wiki-MetaSemantik is simpler than the Persian Wikipedia Ontology QE due to it capture the most significant terms from a concept graph using network properties only.

*Ranking Suggestions*

For all QE methods used in this paper, the postprocessing after QE terms are found as follows: ignore the AND/OR operators and delete stopwords from synonyms list. After that generate QE terms by taking related keywords per user query term and in FCFS order. For example: for two terms basic query "*adolescent and alcoholism*", if |QE terms|=2 then pick 1 synonym from "*adolescent*" and 1 synonym from "*alcoholism*"; if |QE terms|=3 then pick 2 synonyms from "*adolescent*" and 1 synonym from "*alcoholism*". For 3 terms basic query "*Java and applet and programming*" , if |QE terms|=2 then pick 1 synonym from "*Java*" and 1 synonym from "*applet*" only; if |QE terms|=3 then pick 1 synonym from each word.

```
MetaPRF: Generate ℜ
Input:
    user_query    : a user query
    |QE_terms|    : number of terms in QE
    QE_degree(g_concept(qc_x)), QE_closeness(g_concept(qc_x)), QE_pageRank(g_concept(qc_x)),
                  QE_WordNet(g_concept(qc_x)), QE_WikiSynonyms(g_concept(qc_x)),
                  QE_mobyThesaurus(g_concept(qc_x))
                  : list of candidates of QE terms in decrease order as a result of computing
                    {degree|closeness|pageRank|WordNet|WikiSynonyms |mobyThesaurus} from
                    a Wikipedia graph derived from a query concept qc_x given user_query
Output:
    ℜ : a web search results list of a meta-search engine (MSE)

1. Define number of query expansion terms, m = |QE_terms|.
2. DO query rewriting:
   Expanded_query = user_query + top_m {QE_degree(g_concept(qc_x))| QE_closeness(g_concept(qc_x))|
   QE_pageRank(g_concept(qc_x)) | QE_WordNet(g_concept(qc_x))| QE_WikiSynonyms(g_concept(qc_x)) |
   QE_mobyThesaurus(g_concept(qc_x)) }
3. Fetch each expanded query in Step 2 to each component engines separately.
4. FOR each component engines SE_i, i=1,...,n where n is total number of component engines:
   4.1. Set up confidence values for component engines.
        // e.g. SE_conf_i = 30 for Google, SE_conf_i = 25 for Lycos, SE_conf_i = 20 for Bing, SE_conf_i =
        15 for Ask.com, SE_conf_i = 10 for Exalead.
   4.2. Set up weights for 6 knowledge sets.
        // e.g. w_degree = 30, w_closeness = 20, w_pageRank = 20, w_WordNet = 10,
        w_WikiSynonyms = 10, w_mobyThesaurus = 10
   4.3. Set up weight values of each component engines:
        w_SE_degree = w_degree *SE_conf_i / 100
        w_SE_closeness = w_ closeness *SE_conf_i / 100
        w_SE_pageRank = w_ pageRank *SE_conf_i / 100
        w_SE_WordNet = w_ WordNet *SE_conf_i / 100
        w_SE_WikiSynonyms = w_ WikiSynonyms *SE_conf_i / 100
        w_SE_mobyThesaurus = w_ mobyThesaurus *SE_conf_i / 100
5. Create an MSE using top-200 search results from component engines. (See D. Puspitaningrum et
   al., "The Analysis of Rank Fusion Techniques to Improve Query Relevance", (2015)).
```

Fig. 4. The Meta-PRF algorithm: a gold standard algorithm using data fusion

TABLE II
PERFORMANCE OF AVERAGE RUNNING TIME PER QUERY IN INDIVIDUAL SEARCH ENGINES (IN SECONDS). THE SET UP FOR WIKI-METASEMANTIK (DEGREE-CLOSENESS-PAGERANK) IS (20-30-20)

| QE Method | \|QE terms\| = 2 | \|QE terms\| = 3 |
|---|---|---|
| \|User_query\| = 2 terms | | |
| a. Wiki-MetaSemantik | 0.473 | 0.328 |
| b. Persian Ontology | 33.283 | 36.160 |
| \|User_query\| = 3 terms | | |
| a. Wiki-MetaSemantik | 0.249 | 0.327 |
| b. Persian Ontology | 70.030 | 53.749 |

*Evaluation Utility: Automatic Judgement*

Instead of asking the user to identify relevant documents, we simply assume that the top-ranked documents are relevant (*pseudo-relevance feedback*). To do automatic relevance judgement, we compare search results over a meta-search engine or an individual search engine against gold standard, viz. the top-*k* search results of the Meta-PRF algorithm, *k*={3,5,10,20,50} with |QE terms|=10. We choose the top-*200* retrieved documents of each component engines to be merged as an MSE search results because they are in decrease ordered by relevance and that they are the most probable viewed documents by user. We define |QE terms|=10 under assumption that the top-*10* terms of each knowledge sources (WordNet, WikiSynonyms, Moby thesaurus, degree, closeness, pageRank) are higly related terms and they capture well concept or semantic of user query. Following (Fig. 4) is an algorithm for creating the gold standard, viz. the pseudo-relevant dataset. The meta-PRF algorithm takes benefits of data fusion: combine advantages of each component engines. The MSE algorithm viz. Weighted Borda Fuse (or WBF) is as in [18][6]. As inputs we take only synonyms of user query from online thesauruses (WordNet, WikiSynonyms, Moby Thesaurus), as well as synonyms and related terms from Wikipedia by computing ontology graph properties (degree, closeness, pageRank) as in Step 7 in Section 3.

As evaluation criteria we use precision, success and runtime, denoted by P@*x*, S@*x* and time. P@*x* denotes precision of the *x* highest ranked documents with $x \in \{5,10,20,50\}$, and is defined as the average percentage of the first *x* retrieved documents that is relevant with the gold standard, averaged over all documents. S@*x* denotes success of the *x* highest ranked documents with $x \in \{5,10,20,50\}$.

*Evaluation Utility: Human Judgement*

For ground truth, we use Cohen's kappa coefficient to measure the reliability of scoring diagnosis by two human judges. For each query, we take top-*10* retrieved documents from each QE methods to be scored either 0, 1, or 2 where 0=("*not relevant*"), 1=("*partially relevant*"), and 2=("*relevant*"). Then the Cohen's kappa coefficient measures agreement between judges on the same objects and subtracting out agreement due to chance. The kappa coefficient, or $\kappa$, has value around [0,1]. The closer the $\kappa$ coefficient to 1 the more agree the two parties. Once $\kappa$ coefficient shows strong agreement of 2 judges, quality of query expansion system is measured using Normalized DCG (NDCG).

## C. *Results and Discussion*

Table II shows scalability of Wiki-MetaSemantik. Time starts once an ontology graph is created. Persian Wikipedia Ontology QE is so slow fast due to it involves vector processing with 4 times ontology matrix multiplication where maximal index of query vector and ontology matrix is equal to its total nodes. In other hand, Wiki-MetaSemantik works very fast because it computes degree, closeness, and pageRank respectively at once thus time efficient. From Table II, the ratio of running time between Wiki-MetaSemantik vs Persian Wikipedia Ontology QE is ranging between 1:70 and 1:281 for |QE terms|=2 and 1:110 and 1:164 for |QE terms|=3, with the first two is for |User_query|=2 and the latter two is for |User_query|=3. Therefore Wiki-MetaSemantik is very potential to be implemented online.

Parameters tuning in Wiki-MetaSemantik should be treated carefully. We did experiments with different weights of graph properties (degree-closeness-PageRank) viz. (30-20-20), (20-30-20), (20-20-30), and found that (20-30-20) is the best parameters set up (see Fig. 5). This shows the importance of closeness weight parameter, followed by degree weight and pageRank weight. Giving more weight to a graph properties when creating an MSE means we trust more on QE term candidates from its graph properties. About pageRank, giving more weight to pageRank is not too helpful (Fig. 5(c)) since the graph is more like a tree than a cyclomatic graph. Furthermore, the higher the closeness score to a knowledge concept, the more important the node (a QE term candidate). The closeness function is almost similar to degree, that is finding an important node, but closeness is more powerful because it consider the minimum length of path from the concept to other nodes.

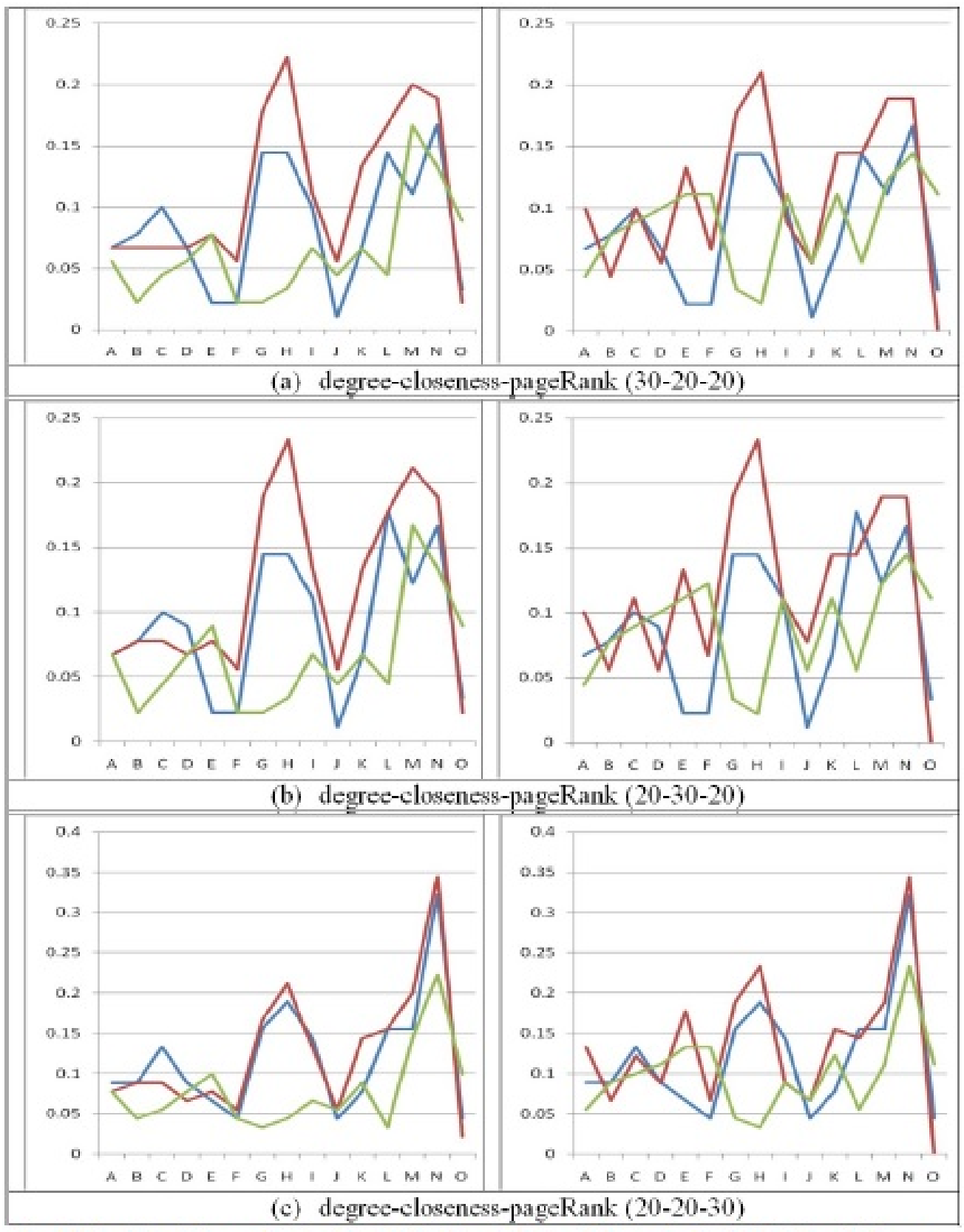

A: MSE (Google-Lycos-Bing), B: MSE (Google-Lycos-Ask.com), C: MSE (Google-Lycos-Exalead), D: MSE (Google-Bing-Ask), E: MSE (Google-Bing-Exalead), F: MSE (Google-Ask-Exalead), G: MSE (Lycos-Bing-Ask), H: MSE (Lycos-Bing-Exalead), I: MSE (Lycos-Ask-Exalead), J: MSE (Bing-Ask-Exalead), K: Google, L: Bing, M: Lycos, N: Exalead, O: Ask.com

No query expansion (blue), Wiki-MetaSemantik (red), Wikipedia Persian Ontology (green)

Fig. 5. P@3 of Wiki-MetaSemantik vs Persian Ontology with |QE terms|=2 (left) and 3 (right)

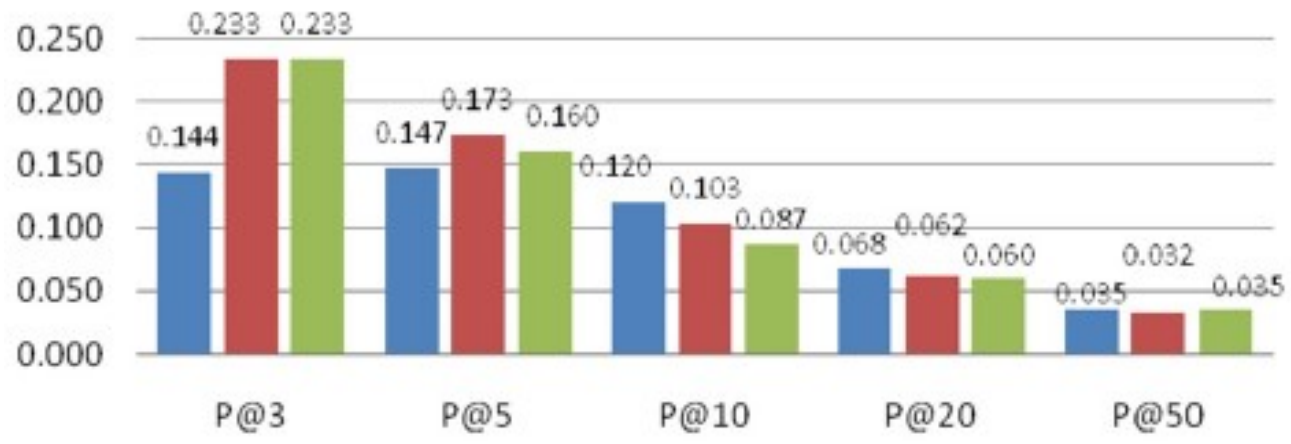

Fig. 6. Precisions of Wiki-MetaSemantik (degree-closeness-pageRank=(20-30-20)). Red line is |QE terms|=2, green is |QE terms|=3, blue is without QE. All is running on MSE (Lycos-Bing-Exalead).

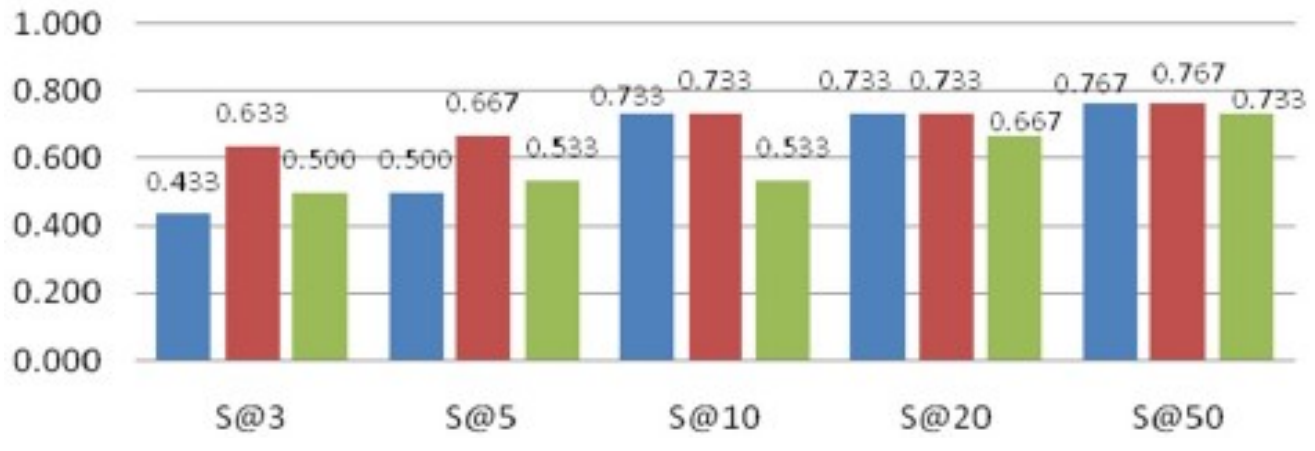

Fig. 7. Successes of Wiki-MetaSemantik (degree-closeness-pageRank=(20-30-20)). Red line is |QE terms|=2, green is |QE terms|=3, blue is without QE. All is running on MSE (Lycos-Bing-Exalead).

| Wiki-Meta Semantik | Persian Ontology | Wikisynonyms | WordNet | Moby |
|---|---|---|---|---|
| \|QE terms\| = 2 | | | | |
| User query = "*adolescent and alcoholism*" | | | | |
| alcoholic beverage | pejorative alcohol | adolescence alcoholic | stripling alcohol | juvenal drug |
| QE rewriting for gold standard (meta-PRF) with \|QE terms\|=10:<br>degree<br>(adolescent and alcoholism) dmoz stereotype public health disability-adjusted life year cocaine addiction ethanol<br>Closeness<br>(adolescent and alcoholism) dmoz stereotype cocaine addiction public health disability-adjusted life year ethanol<br>PageRank<br>(adolescent and alcoholism) self-medication alcohol withdrawal syndrome addictive personality benzodiazepine physical dependence addiction | | | | |

Fig. 8. QE terms for query "*adolescent and alcoholism*". The gold standard (meta-PRF) consists of QE terms from MSE(Lycos-Bing-Exalead).

TABLE III
WIKI-METASEMANTIK PERFORMANCE (HUMAN JUDGEMENT) OF USER QUERY "ADOLESCENT AND ALCOHOLISM" WITH SET UP FOR GRAPH PROPERTIES (DEGREE-CLOSENESS-PAGERANK)=(20-30-20)

| \|QE terms\| | $NDCG_3$ | $NDCG_5$ | $NDCG_7$ | $NDCG_{10}$ |
|---|---|---|---|---|
| 2 | 1.00 | 1.00 | 1.00 | 1.00 |
| 3 | 0.81 | 0.73 | 0.83 | 0.90 |

Using the same set up of graph properties, Fig. 6 and Fig. 7 indicate that Wiki-MetaSemantik shows improvement in relevance of retrieved documents at top-*5* documents and then the relevance decreases after that. This means that the Wiki-MetaSemantik is working well on the locations of the top most viewed documents by search engine users. By Fig. 6 the improvement ratio scores for precisions are: (1:1.62:1.62) and (1:1.18:1.09) which are from P@3 and P@5, and by Fig. 9 the improvement ratio scores for successes are: (1:1.46:1.16) and (1:1.33:1.07) which are from S@3 and S@5. This is a significant improvement, especially in the top-*3* retrieved documents. Therefore Wiki-MetaSemantik helps user find relevant documents effectively.

Fig. 8 shows an example of gold standard QE terms against QE terms from several QE methods. The $m$=10 in Fig. 4 means the gold standard has 10 QE terms taken from top-*10* of BC list of Intersection_set#1, Intersection_set#2, and Intersection_set#3 (see Section 3.2 Step 7). Fig. 8 shows the 10 QE terms in gold standard are qualified because it shows highly related terms with user query.

Fig. 9 shows some examples of QE of our total 30 multi-domain queries. All the terms in the figure are those suggested by each QE methods. It shows that Wiki-MetaSemantik is very good in capturing semantic relatedness. It retrieves highly related QE terms because Wiki-MetaSemantik uses only nodes from best ontology (or best knowledge concept), and filter the nodes using graph properties of (degree, closeness, and pageRank) to produce qualified QE terms.

Furthermore, we prove quality of our Wiki-MetaSemantik search results by random sampling multi domain queries and ask 2 judges for relevance. We found that the $\kappa$ coefficient is ranged from 0.512 to 1, which shows fair to good agreement between the judges. Table III shows average of NDCG scores of two human judges at top-$k$ search results documents ($k$={3,5,7,10}) for Wiki-MetaSemantik on |QE terms|={2,3}. From the figure, Wiki-MetaSemantik with |QE terms|=2, even

shows an ideal ranking until top-*10* of retrieved documents. More QE terms can lead to bias against user query.

In general, Wiki-MetaSemantik shows significant improvement of performance (Fig. 6 and Fig. 7). Wiki-MetaSemantik works better while implemented in an MSE rather than in an individual search engine (Fig. 5) because an MSE combines all advantages of its component engines. Fine tuning of graph properties' weights influences its performance with order of importance as follows: closeness>degree> pageRank. The structure of Wikipedia derived ontology graph influences the pageRank performance due to pageRank good in a cyclomatic graph rather than a tree graph. Allowing indirect links of Wikipedia pages, the minimum length of path (closeness) from user query concept is more important than the number of outlinks a node has (degree). This simplicity as well as an idea of generating an ontology graph on the fly make it suitable for multi domain queries thus we do not have to depend on limited ontologies available in Wikipedia. Also since it works very fast against other Wikipedia query expansion baselines (Table II) thus Wiki-MetaSemantik is potential to be implemented online.

| Wiki-Meta Semantik | Persian Ontology | Wikisynonyms | WordNet | Moby |
|---|---|---|---|---|
| A. \|QE terms\| = 2 | | | | |
| User query = "*adolescent and alcoholism*" | | | | |
| alcoholic beverage | pejorative alcohol | adolescence alcoholic | stripling alcohol | juvenal drug |
| User query = "*programming or algorithm*" | | | | |
| coding theory | ***adaptive-additive entropy*** | computer algorithmics | ***scheduling*** algorithmic | ***mapping formula*** |
| User query = "*geographical or stroke or incidence*" | | | | |
| public health | hypercholesterolemia ***cincinnati*** | geography cerebrovascular | geographic shot | ***science*** stress |
| B. \|QE terms\| = 3 | | | | |
| User query = "*adolescent and alcoholism*" | | | | |
| alcoholic beverage psychological | pejorative alcohol ***cardiovascular*** | adolescence teenage alcoholic | stripling teenage alcohol | young teener drug |
| User query = "*programming or algorithm*" | | | | |
| coding theory information | ***adaptive-additive entropy binary*** | computer software algorithmics | ***scheduling*** programing algorithmic | ***document rule formula*** |
| User query = "*geographical or stroke or incidence*" | | | | |
| public health methamphetamine | hypercholesterolemia ***cincinnati scale*** | geography cerebrovascular ***angle*** | geographic shot ***relative*** | ***science*** stress ***routineness*** |

Fig. 9. Examples of QE terms. In *bold italics* are bad terms.

## V. CONCLUSIONS

We have been proposed Wiki-MetaSemantik, a query expansion technique using an ontology graph derived from Wikipedia that captures semantic relatedness very well. The results show that by using Wiki-MetaSemantik, relevance of the retrieval system is improved as well as time efficient.

Possible limitation of Wiki-MetaSemantik is the quality of QE terms it produced may fall depending on the quality of Wikipedia pages and the links in these pages. Therefore for future work, we suggest the tight integration of Wiki-MetaSemantik and online thesauruses to improve the success of ontology-based query expansion terms over the plain (only synonyms) user query terms to cover up lack of knowledge due to inexistence of Wikipedia pages on certain topics.